%% file: main.tex
\documentclass[sigconf]{acmart}

\usepackage{enumitem}
\usepackage{multirow}
\usepackage{booktabs}
\usepackage{array}
\usepackage{subfigure}
\usepackage{float}
\usepackage{stfloats}
\usepackage{balance}
\usepackage{marvosym}

\newenvironment{shrinkeq}[1] 
{ 
  \bgroup
  \addtolength\abovedisplayshortskip{#1}
  \addtolength\abovedisplayskip{#1}
  \addtolength\belowdisplayshortskip{#1}
  \addtolength\belowdisplayskip{#1}
}
{\egroup\ignorespacesafterend}

\AtBeginDocument{%
  }

\setcopyright{acmlicensed}
\copyrightyear{2018}
\acmYear{2018}
\acmDOI{XXXXXXX.XXXXXXX}
\acmConference[Conference acronym 'XX]{Make sure to enter the correct
  conference title from your rights confirmation email}{June 03--05,
  2018}{Woodstock, NY}
\acmISBN{978-1-4503-XXXX-X/2018/06}

\begin{document}

\title{FuXi-Linear: Unleashing the Power of Linear Attention in Long-term Time-aware Sequential Recommendation}

\author{Yufei Ye}
\email{aboluo2003@mail.ustc.edu.cn}
\affiliation{%
  \institution{University of Science and Technology
of China}
  \city{Hefei}
  \country{China}
}

\author{Wei Guo\textsuperscript{\Letter}}
\email{guowei67@huawei.com}
\affiliation{
  \institution{Huawei Technologies}
   \city{Shanghai}
  \country{China}
}

\author{Hao Wang\textsuperscript{\Letter}}
\email{wanghao3@ustc.edu.cn}
\affiliation{%
  \institution{University of Science and Technology
of China}
  \city{Hefei}
  \country{China}
}

\author{Luankang Zhang}
\email{zhanglk5@mail.ustc.edu.cn}
\affiliation{%
  \institution{University of Science and Technology
of China}
  \city{Hefei}
  \country{China}
}

\author{Heng Chang}
\email{heng.chang@huawei.com}
\affiliation{
  \institution{Huawei Technologies}
   \city{Beijing}
  \country{China}
}

\author{Hong Zhu}
\email{zhuhong8@huawei.com}
\affiliation{
  \institution{Huawei Technologies}
   \city{Shenzhen}
  \country{China}
}

\author{Yuyang Ye}
\email{yeyuyang@mail.ustc.edu.cn}
\affiliation{%
  \institution{University of Science and Technology
of China}
  \city{Hefei}
  \country{China}
}

\author{Yong Liu}
\email{liu.yong6@huawei.com}
\affiliation{
  \institution{Huawei Technologies}
   \city{Shanghai}
  \country{China}
}

\author{Defu Lian\textsuperscript{\Letter}}
\email{liandefu@ustc.edu.cn}
\author{Enhong Chen}
\email{cheneh@ustc.edu.cn}
\affiliation{%
  \institution{University of Science and Technology
of China}
  \city{Hefei}
  \country{China}
}


\renewcommand{\shortauthors}{Yufei Ye et al.}

\begin{abstract}
    Modern recommendation systems primarily rely on attention mechanisms with quadratic complexity, which limits their ability to handle long user sequences and slows down inference. While linear attention is a promising alternative, existing research faces three critical challenges: (1) temporal signals are often overlooked or integrated via naive coupling that causes mutual interference between temporal and semantic signals while neglecting behavioral periodicity; (2) insufficient positional information provided by existing linear frameworks; and (3) a primary focus on short sequences and shallow architectures.
    To address these issues, we propose FuXi-Linear, a linear-complexity model designed for efficient long-sequence recommendation. Our approach introduces two key components: (1) a Temporal Retention Channel that independently computes periodic attention weights using temporal data, preventing crosstalk between temporal and semantic signals; (2) a Linear Positional Channel that integrates positional information through learnable kernels within linear complexity. Moreover, we demonstrate that FuXi-Linear exhibits a robust power-law scaling property at a thousand-length scale, a characteristic largely unexplored in prior linear recommendation studies. Extensive experiments on sequences of several thousand tokens demonstrate that FuXi-Linear outperforms state-of-the-art models in recommendation quality, while achieving up to 10$\times$ speedup in the prefill stage and up to 21$\times$ speedup in the decode stage compared to competitive baselines. Our code has been released in a public repository \textcolor{blue}{\url{https://github.com/USTC-StarTeam/fuxi-linear}}.

\end{abstract}



\received{20 February 2007}
\received[revised]{12 March 2009}
\received[accepted]{5 June 2009}

\settopmatter{printacmref=false} 
\setcopyright{none} 
\renewcommand\footnotetextcopyrightpermission[1]{} 

\maketitle

\input{chapters/arxiv_1_introduction}

\input{chapters/arxiv_2_related_work}

\input{chapters/arxiv_3_method}

\input{chapters/4_analysis}

\input{chapters/5_experiments}

\input{chapters/6_conclusions}

\bibliographystyle{ACM-Reference-Format}
\bibliography{main}

\input{chapters/7_appendix}

\end{document}

%% file: chapters/arxiv_1_introduction.tex
\section{Introduction}

Recommender Systems (RS) have become indispensable in alleviating information overload, where sequence modeling serves as the fundamental backbone. From traditional two-stage modeling frameworks \cite{pi2020search, zhou2019deep, chang2023twin, si2024twin,wang2025enhancing,zhou2025multi} to the emerging paradigm of generative recommendation \cite{zhai2024actions, yang2025sparse, deng2025onerec, han2025mtgr, wang2025generative, wang2025dlf, zhang2025killing}, the ability to capture evolving user preferences from historical interaction sequences remains a pivotal component for predictive accuracy.

The most commonly used method for sequence modeling is softmax-based attention \cite{kang2018self,zhai2024actions,yinfeature}. However, its quadratic computational complexity $O(n^2)$ with respect to sequence length $n$ imposes a prohibitive bottleneck in long-term sequence scenarios, severely limiting both the exploitable sequence length and inference throughput. While two-stage architectures utilizing target attention \cite{pi2020search} offer improved efficiency, they inevitably sacrifice rich contextual dependencies within the sequence. In modern industrial RS, models are increasingly required to process user histories exceeding $10^4$ interactions \cite{guan2025make, si2024twin, huang2024chemeval}, making high-capacity yet efficient modeling a necessity. Recently, linear attention has emerged as a promising alternative in NLP \cite{sun2023retentive, dao2024transformers, team2025kimi, yu2025thought, lv2025costeer}, achieving linear training complexity and constant-time inference while maintaining performance comparable to Transformers.

Despite their potential, directly applying linear attention to RS encounters notable limitations given the unique temporal and positional sensitivities of recommendation data \cite{ye2025fuxib}. First, existing linear models often overlook interaction timestamps \cite{yue2024linear, yang2024ttt4rec} or couple temporal signals as scalar scales into semantic attention weights~\cite{fan2025tim4rec, xiao2025ss4rec}. Such an approach causes semantic signals and temporal signals to interfere with each other, severely compromising the model's utilization of temporal information. Furthermore, traditional methods for integrating temporal information \cite{zhai2024actions, ye2025fuxi} violate the recurrent form of linear attention, causing complexity to revert to $O(n^2)$. Notably, current models fail to explicitly account for periodicity, which is useful for capturing long-term user behaviors. Second, although the attenuation of the linear models~\cite{sun2023retentive, gu2023mamba} at each step can provide certain positional information, it lacks the granularity to distinguish between adjacent positions. Powerful alternatives like Relative Position Encoding (RPE) \cite{raffel2020exploring, press2021train} are incompatible with linear recurrence, and recent studies suggest that Rotary Positional Embedding \cite{su2024roformer} underperforms compared to RPE in sequential recommendation tasks \cite{zhai2024actions}. Furthermore, most existing explorations of linear attention in sequential recommendation~\cite{wang2024echomamba4rec,fan2025tim4rec, xiao2025ss4rec} remain constrained to short-sequence settings (e.g., $\le 100$ interactions) and shallow architectures (e.g., 1–2 layers), leaving their scalability and capacity in ultra-long sequence scenarios largely unverified.

To bridge these gaps, we propose FuXi-Linear, a novel linear-complexity architecture designed for long-sequence recommendation. We introduce the Temporal Retention Channel, which utilizes period-aware queries and keys generated solely from timestamps. This design avoids mutual interference between temporal and semantic signals and effectively captures periodic patterns without sacrificing linear efficiency. To supplement positional signals, we develop a learnable kernel-based mechanism that approximates the efficacy of RPE while maintaining a recurrent form. Moreover, unlike existing linear recommendation models that are largely confined to shallow architectures and short sequences, we demonstrate that FuXi-Linear exhibits a robust power-law scaling property at a thousand-length scale. Extensive experiments on three real-world datasets demonstrate that FuXi-Linear significantly outperforms state-of-the-art baselines. Remarkably, compared to the most competitive baselines, FuXi-Linear achieves up to 10$\times$ speedup in the prefill stage and up to 21$\times$ speedup during decoding.
Our main contributions are summarized as follows:
\begin{itemize}[leftmargin=*,align=left]
    \item We propose FuXi-Linear, a novel sequential recommendation model that addresses the efficiency-effectiveness dilemma in long sequence modeling via a refined linear attention mechanism.
    \item We design the Temporal Retention Channel, which effectively captures periodic temporal signals and avoids semantic interference while adhering to linear complexity.
    \item We introduce a learnable kernel function for positional modeling, which provides the expressive power of relative position encoding while maintaining linear complexity.
    \item We conduct extensive experiments on three real-world benchmarks, demonstrating that FuXi-Linear achieves state-of-the-art performance. As sequence length grows, it provides up to 21$\times$ faster inference than competitive baselines.
\end{itemize}

%% file: chapters/arxiv_2_related_work.tex
\section{Related Work}
\label{sec:related_work}

\subsection{Linear Attention}
In the field of language models, there have been numerous attempts at linear attention mechanisms, including approaches based on linear RNNs \cite{sun2023retentive, yang2023gated, zhang2024gated, du2025mom, yu2025thought}, State Space Models \cite{gu2023mamba, dao2024transformers}, TTT \cite{sun2024learning, behrouz2024titans, xie2025breaking}, and Delta Rule methods \cite{yang2024parallelizing, yang2024gated, team2025kimi}. However, relatively fewer works have explored these techniques in recommendation systems. LinRec \cite{liu2023linrec} achieves linearization by first computing the KV product, while LRURec \cite{yue2024linear} and RecBLR \cite{liu2024behavior} implement linearization through Linear Recurrent Units. Approaches like Mamba4Rec \cite{liu2024mamba4rec}, SSD4Rec \cite{qu2024ssd4rec}, and TTT4Rec \cite{yang2024ttt4rec} leverage their respective linear attention mechanisms to model user interests. More recently, EchoMamba4Rec \cite{wang2024echomamba4rec} and SIGMA \cite{liu2025sigma} have employed bidirectional Mamba architectures for user sequence modeling, incorporating additional techniques to enhance recommendation performance. However, these studies on sequential recommendation models with linear attention are primarily based on small models handling short sequences, and they lack sufficient temporal positional information.

\subsection{Time-aware Sequential Recommendation}

A key difference between item sequences and text is the inclusion of temporal information \cite{zhai2024actions, shen2025genki, tong2024mdap,wang2024denoising, gu-etal-2025-rapid}. TiSASRec \cite{li2020time} integrates temporal information by generating embedding vectors based on relative time intervals and adding them to the key vectors. MOJITO \cite{tran2023attention} employs a Gaussian mixture model to merge temporal attention weights with item attention weights. SS4Rec \cite{xiao2025ss4rec} utilizes time intervals to perform discretization with the S5 model. TiM4Rec \cite{fan2025tim4rec} introduces the Time-aware Structured Masked Matrix and incorporates it into the SSD model. TAT4SRec \cite{zhang2023time} utilizes bucketing and window function weighting, encoding temporal information through an Encoder, and then integrating it into a Decoder. HSTU \cite{zhai2024actions} adds temporal information using bucketed relative attention bias. FuXi-$\alpha$ \cite{ye2025fuxi} utilizes this bias as a separate attention map, while FuXi-$\beta$\cite{ye2025fuxib} employs a continuous function to avoid the hardware inefficiency of the bucketing process.

However, existing approaches either require quadratic time complexity or interfere with semantic signals. Moreover, they fail to account for periodicity.

\subsection{Positional Embeddings}
In Transformers, positional information is integrated using three main approaches: absolute positional encoding, relative positional encoding, and rotary positional encoding. Absolute positional encoding represents position information as vectors added directly to embeddings. The original Transformer utilized sinusoidal positional embeddings \cite{vaswani2017attention}, while BERT \cite{devlin2019bert} allocates learnable positional vectors for each position. Relative positional encoding, on the other hand, incorporates relative information within the attention layer. In works such as \cite{shaw2018self}, Transformer-XL \cite{dai2019transformer}, and DeBERTa \cite{he2020deberta}, modifications are made to the key or value matrices based on relative positional information. T5 \cite{raffel2020exploring} and Alibi \cite{press2021train} generate bias terms for attention weights through relative positional information. Recent work \cite{ye2025fuxi,ye2025fuxib} regard this bias term as an attention map to model positional information. Rotary positional encoding \cite{su2024roformer} involves rotating the dimensions of the query and key in groups, thereby incorporating both absolute and relative positional information during attention computation. XPos \cite{sun2022length} extends RoPE by introducing decay terms to improve extrapolation. 

However, absolute positional embeddings lack relative detail, and rotary positional embeddings underperform in recommendation tasks compared to the costly relative positional embeddings~\cite{zhai2024actions}, which require a quadratic time complexity.

%% file: chapters/arxiv_3_method.tex
\section{Problem Statement}
In time-aware sequential recommendation, the goal is to predict the next item a user will interact with based on their past interactions \cite{kang2018self,yu2025thought,huang2025selfaug}. Along with item and position information, timestamps of interactions are crucial. 

Formally, given a set of users $\mathcal{U}$ and items $\mathcal{I}$, each user $u \in \mathcal{U}$ has an interaction sequence:
\[
\mathcal{S}_u = \left[\left(i_1^{(u)}, t_1^{(u)}\right), \left(i_2^{(u)}, t_2^{(u)}\right), \ldots, \left(i_{n_u}^{(u)}, t_{n_u}^{(u)}\right)\right],
\]
where $t_j^{(u)}$ denotes the timestamp when the user interacted with item $i_j^{(u)}$, $n_u$ represents the length of the sequence for user $u$ and items are ordered chronologically.

The task is to predict the next item $i_{n_u+1}^{(u)}$ by estimating the probability distribution over $\mathcal{I}$: $P(i_{n_u+1}^{(u)} = i \mid \mathcal{S}_u)$ for $i \in \mathcal{I}$. During training, the model learns to predict the next item $i_{j+1}^{(u)}$ for each prefix of the sequence $\mathcal{S}_u$, with the target sequence being $[i_2^{(u)}, i_3^{(u)}, \ldots, i_{n_u+1}^{(u)}]$ \cite{kang2018self}.

\section{Methodology}

\begin{figure*}
    \centering
    \includegraphics[width=0.9\linewidth]{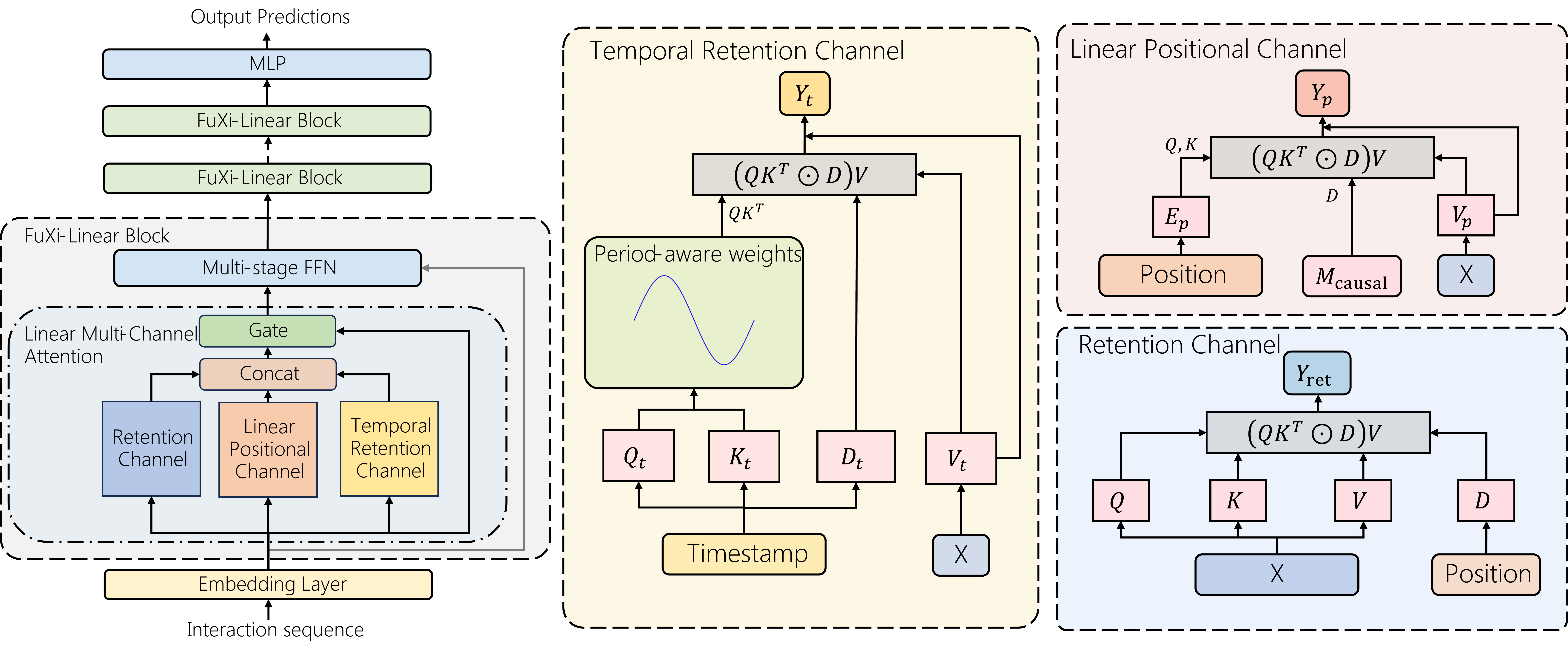}
    \caption{The overall architecture of FuXi-Linear.}
    \label{fig:structure-overview}
    \vspace{-10pt}
\end{figure*}

The overall architecture of our model is illustrated in Figure \ref{fig:structure-overview}, primarily composed of several stacked FuXi-Linear blocks. In each FuXi-Linear block, the input is transformed through three separate channels, and the results are concatenated and passed through gating mechanisms. This is followed by merging through the Multi-stage Feed-forward Network (MFFN). Each channel supports three computational forms: recurrent, parallel, and chunkwise recurrent. The \textbf{recurrent form} is employed for $O(1)$ inference, the \textbf{parallel form} is suitable for parallel training, and the \textbf{chunkwise recurrent form} is used for parallel training with linear complexity.

\subsection{Embedding Layer}

Each user's interaction sequence is transformed into a fixed-length sequence of size $n$ via truncation or padding prior to the embedding layer. Sequences shorter than $n$ are extended by appending a special padding item. Within the embedding layer, every item $i \in \mathcal{I}$ is projected into a $d$-dimensional dense vector representation through a trainable embedding matrix $E \in \mathbb{R}^{\mathcal{I} \times d}$, where $d$ denotes the dimensionality of the latent vector space. Subsequently, a learnable absolute positional embedding is added to each non-padding item.

\subsection{Retention Channel}

In this channel, we use a simple linear attention, the retention mechanism\cite{sun2023retentive}, to replace the full attention which has quadratic complexity, thus providing semantic information.
\newline \textbf{Parallel Form}. Suppose the input for the current layer is $X\in \mathbb{R}^{n\times d}$. We then compute the query, key, and value matrices as follows:
$$
Q = \phi(XW_{q}), \quad K = \phi(XW_k), \quad V = \phi(XW_v)
$$
where $\phi$ denotes the SiLU activation function. Next, query-key attention is calculated and element-wise multiplied by a matrix $D$, which is computed based on positional differences:
\begin{align}
    \text{Retention}(Q, K, V, D)=(QK^T\odot D) V \\
    D_{i,j} = \begin{cases}
      \gamma^{i-j}, & i\geq j \\
      0, & \text{otherwise}
    \end{cases}
\end{align}
Here, $\gamma$ is a learnable parameter. We extend this approach to multiple heads, where the query, key, and value matrices for each head are denoted as $(Q_1, K_1, V_1), \cdots, (Q_h, K_h, V_h)$. The learnable parameter for the $i$-th head is $\gamma_i$, and the corresponding matrix $D$ is denoted as $D_i$. The output of the retention channel, $Y_{\text{ret}}$, is given by:
\begin{align}
    \text{head}_i &= \text{Retention}(Q_i, K_i, V_i, D_i) \\
    Y_{\text{ret}} &= \text{Norm}(\text{Concat}(\text{head}_1, \cdots, \text{head}_h))
\end{align}
\newline \textbf {Recurrent Form}. Suppose the input at step $n$ is $x_n \in \mathbb{R}^{1\times d}$. Using the same methodology, we compute the query, key, and value vectors for the $i$-th head as $q_n^{(i)}, k_n^{(i)} \in \mathbb{R}^{1\times d_k}, v_n^{(i)}\in \mathbb{R}^{1\times d_v}$. Let $S_n^{(i)} \in \mathbb{R}^{d_{k}\times d_{v}}$ denote the hidden state of the $i$-th head, with the following equation:
\begin{align}
S_n^{(i)} = \gamma_{i} S_{n-1}^{(i)} + (k_n^{(i)})^T v_n^{(i)}
\end{align}
Thus, the output for the $i$-th head is:
\begin{align}
\text{head}_i = q_{n}^{(i)} S_{n}^{(i)}
\end{align}

\subsection{Linear Positional Channel}

We aim to compute the output of the positional channel, represented by $\sum_{i = 1}^{n} f(n - i) v_{i}$, where $f$ is a type of relative position encoding method, such as T5-style relative attention bias \cite{raffel2020exploring}, Alibi \cite{press2021train}, and others. However, these methods typically require quadratic computational complexity. To mitigate this, we propose modeling $f$  using learnable kernel functions, enabling a recurrent formulation.

Relaxing the constraint of translation invariance, we approximate $f(x - y)$ with a kernel function:
\begin{align}
f(x - y) \approx g(x, y) = \mathbf{k}^T(x) \mathbf{k}(y)
\end{align}
where $\mathbf{k}$ is a mapping function that projects positions $x$ and $y$ into a $d_p$-dimensional vector space. Consequently, the output of the positional channel can be reformulated as:
\begin{align}
(y_{p})_n &= \sum_{i = 1}^n g(n, i) v_i \\
&= \mathbf{k}(n) \sum_{i = 1}^n \mathbf{k}(i)^T v_i.
\end{align}
\textbf{Recurrent Form} We first obtain the projection $(v_p)_i = x_i W_{p}$ and define the hidden state $(S_p)_n = \sum_{i = 1}^n \mathbf{k}(i)^T v_i \in \mathbb R^{d_p\times d}$ to enable recurrent calculation. For output generation, we enhance the current representation with two learnable parameters $\alpha, \beta \in \mathbb{R}$:
\begin{align}
(S_p)_n &= (S_p)_{n - 1} + \mathbf k(n)^{T} (v_p)_n\\
(y_p)_n &= \alpha \mathbf k(n) (S_p)_n + \beta (v_p)_n
\end{align}
In practice, a learnable embedding matrix $E_p \in \mathbb{R}^{n \times d_p}$ is used to represent the mapping $\mathbf{k}$. 

\textbf{Parallel Form.} Based on the recurrent formulation, the parallel computation for the positional channel can be derived as:
\begin{align}
V_p &= XW_p \\
Y_p &= \left(\alpha(E_p E_p^T \odot M_{\text{causal}}) + \beta I\right)V_p
\end{align}
where $M_{\text{causal}}$ denotes the lower-triangular causal mask, and $I$ is the identity matrix.

\subsection{Temporal Retention Channel}
In SS4Rec \cite{xiao2025ss4rec} and TiM4Rec  \cite{fan2025tim4rec}, temporal information is incorporated by modifying the attention calculation in Mamba, specifically the $QK^T \odot D$ term, where the $D$ matrix is adjusted to encode time-related data. However, this approach causes interference between semantic and temporal information. To address this issue and introduce periodicity, we calculate periodic \( Q_t \) and \( K_t \) using only temporal information, as well as \( D_t \), and compute the output using \( (Q_t K_t^T \odot D_t)V_t \).

To model periodic temporal changes, we represent time dynamics in the complex domain. Assume that, at a given moment, a user’s information is represented by the vector \( v \in \mathbb{R}^{1 \times d} \). After a time interval \( \Delta t \), this vector evolves according to the formula \( v' = e^{\lambda \Delta t}v \), where \( \lambda \in \mathbb{C} \). Writing \( \lambda \) in polar form as \( \lambda = r e^{i\theta} \), the transformation becomes:
\begin{align}
v' = r^{\Delta t} e^{i(\Delta t \theta)} v = r^{\Delta t}\cos(\Delta t \cdot \theta)v + ir^{\Delta t}\sin(\Delta t \cdot \theta)v.
\end{align}
To ensure numerical stability, we require \( 0 < r < 1 \).

Given the extensive temporal range of user data, a single decay factor $\lambda$ is insufficient to capture the multifaceted evolution of interests. We therefore employ a set of $H_t$ parameter pairs $\{(r_h, \theta_h)\}_{h=1}^{H_t}$ to model multi-scale temporal dynamics. However, in long sequences, the accumulated historical context in each head can sometimes overshadow the user's immediate intent. To help the model better identify and prioritize the current representation, we introduce learnable parameters $\alpha_t, \beta_t \in \mathbb{R}^{2H_t}$ to adaptively balance the historical trends and the current state.

Specifically, let $(v_t)_n^{(h)} = x_n W_t^{(h)}$ be the projection of the current interaction $x_n$ onto the $h$-th subspace via the learnable matrix $W_t^{(h)} \in \mathbb{R}^{d \times (d/2H_t)}$. The temporal heads are then formulated as:
\begin{align}
\text{head}_{2h-1} &= \sum_{i=1}^{n} r_h^{t_{n+1}-t_i} \cos((t_{n+1}-t_i) \theta_h) x_i W_t^{(2h-1)} \\
\text{head}_{2h} &= \sum_{i=1}^{n} r_h^{t_{n+1}-t_i} \sin((t_{n+1}-t_i) \theta_h) x_i W_t^{(2h)}
\end{align}
The final output $y$ is obtained by integrating the temporal context with the current state $(v_{t})_n^{(h)}$:
\begin{align}
y = \text{Concat}  [ (\alpha_t)_1 \text{head}_1 + (\beta_t)_1 (v_t)_n^{(1)}, \dots, \nonumber \\
(\alpha_t)_{2H_t} \text{head}_{2H_t} + (\beta_t)_{2H_t} (v_t)_n^{(2H_t)} ]
\end{align}

To derive the recurrent form, consider the following trigonometric identities:

\begin{align}
\sin((t_{n + 1} - t_i) \theta) &= \sin(\theta \cdot t_{n+1}) \cos(\theta \cdot t_i) - \cos(\theta \cdot t_{n+1}) \sin(\theta \cdot t_i) \nonumber \\
&= \begin{pmatrix} \sin(\theta \cdot t_{n + 1}) \\ -\cos (\theta \cdot t_{n + 1}) \end{pmatrix}^T
\begin{pmatrix} \cos (\theta \cdot t_{i}) \\ \sin(\theta \cdot t_{i}) \end{pmatrix}, \\
\cos((t_{n + 1} - t_i) \theta) &= \cos(\theta \cdot t_{n+1}) \cos(\theta \cdot t_i) + \sin(\theta \cdot t_{n+1}) \sin(\theta \cdot t_i) \nonumber \\
&= \begin{pmatrix} \cos(\theta \cdot t_{n + 1}) \\ \sin(\theta \cdot t_{n + 1}) \end{pmatrix}^T
\begin{pmatrix} \cos(\theta \cdot t_{i}) \\ \sin(\theta \cdot t_{i}) \end{pmatrix}.
\end{align}

Let the query and key vectors in the $(2h-1)$-th and $2h$-th heads be defined as follows:

\begin{align}
(q_t)_n^{(2h-1)} &= r_h^{t_{n+1}-t_n} \begin{pmatrix} \sin(\theta_h \cdot t_{n + 1}) \\ -\cos (\theta_h \cdot t_{n + 1}) \end{pmatrix}^T, \\
(q_t)_n^{(2h)} &= r_h^{t_{n+1}-t_n} \begin{pmatrix} \cos(\theta_h \cdot t_{n + 1}) \\ \sin(\theta_h \cdot t_{n + 1}) \end{pmatrix}^T, \\
(k_t)_n^{(2h-1)} &= (k_t)_n^{(2h)} = \begin{pmatrix} \cos(\theta_h \cdot t_{n}) \\ \sin(\theta_h \cdot t_{n}) \end{pmatrix}^T.
\end{align}

Additionally, define $(D_t)^{(2h-1)}_{n,i} = (D_t)^{(2h)}_{n,i} = r_h^{t_{n}-t_i}$ and $(\gamma_t)^{(2h-1)}_n = (\gamma_t)^{(2h)}_n = r_h^{t_{n}-t_{n-1}}$. Using these definitions, two equivalent computation forms can be derived:

\textbf{Recurrent Form} For the $h$-th head, the hidden state $(S_t)_n^{(h)}$ is given by:
\begin{align}
(S_t)_n^{(h)} = \sum_{i=1}^n (D_t)_{n,i}^{(h)} \left[ (k_t)_i^{(h)} \right]^T x_i W_t^{(h)}.
\end{align}
The recurrent update and output computation are:
\begin{align}
(S_t)_n^{(h)} &= (\gamma_t)_n^{(h)} (S_t)_{n-1}^{(h)} + \left[ (k_t)_n^{(h)} \right]^T x_n W_t^{(h)}, \\
\text{head}_h &= (q_t)_n^{(h)} (S_t)_n^{(h)}.
\end{align}
\textbf{Parallel Form} Let the query and key matrices for the $h$-th head be $Q_t^{(h)}, K_t^{(h)} \in \mathbb{R}^{n \times 2}$. The corresponding output is computed as:
\begin{align}
V_t^{(h)} &= XW_t^{(h)} \\
\text{head}_h &= \text{Retention}(Q_t^{(h)}, K_t^{(h)}, V_t^{(h)}, D_t^{(h)}).
\end{align}
To ensure the trigonometric identities hold under 32-bit floating-point operations and to model the user's interests across different time scales, we set \( \theta_k = \frac{2\pi}{B^{b_0 + k}} \). In this formulation, $B$ and $b_0$ are hyperparameters, where both $B$ and $B^{b_0}$ are required to be integers. This ensures the periodicity of the trigonometric functions, $B^{b_0 + k}$, remains an integer. When calculating \( \theta \cdot t \), first take the modulus with respect to the period and then multiply by \( 2\pi \), ensuring the precision limitations of 32-bit floats are satisfied. And \( r_k \) is initialized as \( 2^{-\theta_k /2\pi} \).

\subsection{FuXi-Linear Block}

In the FuXi-Linear Block, the input first passes through the Linear Multi-Channel Attention (LMCA) layer, followed by the Multi-stage Feed-Forward Network (MFFN). Assume that the input to this layer is given by \( X_0 \in \mathbb{R}^{n \times d} \). The output is computed as follows:
\begin{align}
  O &= \text{LMCA}(\text{Norm}(X_0)) \\
  Y &= \text{MFFN}(O, X_0)
\end{align}
In our model, the normalization function utilized is RMSNorm \cite{zhang2019root}.

\subsubsection{Linear Multi-Channel Attention}
Suppose the input to the module is $X \in \mathbb{R}^{n \times d}$. We provide $X$ to three different channels and obtain outputs $Y_{\text{ret}}$, $Y_{p}$, and $Y_{t}$ from each channel, respectively. After normalizing these outputs, we concatenate them and then compute a projection using $X$ to obtain a tensor $U$. The resulting tensor is element-wise multiplied by the concatenated output. This process can be viewed as a gating mechanism to filter out noise. The output of LMCA can be formulated as follows:
\begin{shrinkeq}{-2pt}
    \begin{align}
    \text{LMCA}(X) = \text{Concat}(\text{Norm}(Y_{\text{ret}}), \text{Norm}(Y_{p}), \text{Norm}(Y_{t})) \odot (XW_u)
    \end{align}
\end{shrinkeq}

\subsubsection{Multi-stage Feed-forward Network}

The MFFN consists of two stages, with the first stage merging multi-channel outputs and residual connections, and the second stage facilitating further feature interactions \cite{ye2025fuxi}. Assume the input to this module is represented as \( X \in \mathbb{R}^{n \times 3d} \). The output of the MFFN is given by:
\begin{align}
  Y_{\text{stage1}} &= X W_{0} + X_0 \\
  \tilde Y_{\text{stage1}} &= \text{Norm}(Y_{\text{stage1}}) \\
  \text{MFFN}(X, X_0) &= \left((\tilde Y_{\text{stage1}} W_1) \odot \phi(\tilde Y_{\text{stage1}} W_2)\right) W_3
\end{align}
where \( W_0 \in \mathbb{R}^{3d \times d} \), \( W_1, W_2 \in \mathbb{R}^{d \times d_{\text{FFN}}} \), and \( W_3 \in \mathbb{R}^{d_{\text{FFN}} \times d} \) are learnable parameters. The function \(\phi\) represents the SiLU activation function, and \(\odot\) denotes the element-wise product operation.

\subsection{Chunkwise Recurrent Form}
The complexity of using the parallel form during training remains quadratic in relation to the sequence length, while employing the recurrent form presents challenges in parallel computation. To address the efficiency issues in training, we apply a chunkwise recurrent method \cite{sun2023retentive}. This approach ensures improvements in the limitations of parallel computation within the recurrent form while maintaining low computational complexity.

The parallel computation processes across these three channels can all be abstracted to the expression $Y=(QK^T \odot D)V$, where $Q, K \in \mathbb{R}^{N \times d_k}, V \in \mathbb{R}^{N \times d_v}$, and $D \in \mathbb{R}^{N \times N}$ is a lower triangular matrix. The recurrent computation process can be expressed as
\begin{align}
S_i &= \gamma S_{i-1} + k_i^T v_i \\
y_i &= q_i S_i
\end{align}
For simplicity, we assume that $\gamma$ is a value independent of $i$. Assuming a chunk size of $C$, we define the chunkwise variable $Q_{[t]}^i = q_{tC+i}$, $Q_{[t]} = Q_{[t]}^{1:C} \in \mathbb{R}^{C \times d}$, and similarly define $K_{[t]}, V_{[t]}, Y_{[t]}$. Specifically, the step-wise output and chunk state are derived as:
\begin{align}
    Y_{[t]}^i = Q_{[t]}^i S_{[t]}^i &= \left (\gamma^{i} Q_{[t]}^{i} \right) S_{[t-1]}^C + \sum_{j=1}^i \gamma^{i-j} (Q_{[t]}^i (K_{[t]}^j)^T) V_{[t]}^j \\
    S_{[t]}^C &= \gamma^C S_{[t-1]}^C + \sum_{i=1}^C \gamma^{C-i} (K_{[t]}^i)^T V_{[t]}^i
\end{align}
Let $\tilde{Q}_{[t]}^i = \gamma^i Q_{[t]}^i, D_{\text{in-chunk}} = (d_{i,j})_{C \times C}, \tilde{K}_{[t]}^i = \gamma^{C-i} K_{[t]}^i$, where 
\begin{align}
    d_{ij} = 
    \begin{cases} 
    \gamma^{i-j}, & i \geq j \\ 
    0, & \text{otherwise}
    \end{cases}
\end{align}
The vectorized chunk-level computation is then formulated as:
\begin{align}
Y_{[t]} &= \tilde{Q}_{[t]} S_{[t-1]}^C + \left( Q_{[t]} (K_{[t]})^T \odot D_{\text{in-chunk}} \right) V_{[t]} \\
S_{[t]}^C &= \gamma^C S_{[t-1]}^C + (\tilde K_{[t]})^T V_{[t]}
\end{align}
During training, we perform the above calculations chunk by chunk.

\subsection{Prediction Layer \& Optimization Objective} 

After processing through $L$ layers of the FuXi-Linear Block, the model generates latent representations for each position. Given a user $u$'s interaction history $i_1^{(u)}, \dots, i_n^{(u)}$, the probability of the next item $i_{n+1}$ is predicted using the $n$-th latent representation $x_n$:
\begin{align}
P(i_{n+1} = i_j \mid i_1^{(u)}, \dots, i_n^{(u)}) = \text{softmax}(x_n E)_j
\end{align}
For efficient training, the model is trained autoregressively using a sampled softmax loss with $N$ negative samples \cite{Klenitskiy_2023}.

%% file: chapters/4_analysis.tex
\section{Analysis}\label{Analysis}

\subsection{Complexity Analysis}

We conduct a simplified analysis of the space and time complexity required to compute a user sequence using a single layer of attention. Let us assume the chunk size is \( C \), the sequence length is \( n \), and the embedding size is \( d \).

\begin{itemize}[leftmargin=*,align=left]

  \item \textbf{Space Complexity}. Storing the parameters \( Q, K, V \) for each item requires \( O(nd) \) space. In chunkwise recurrent computation, calculating the key-value product for each chunk requires \( O\left(\frac{n}{C}d^2\right) \) space, while storing the attention map requires \( O(nC) \) space. Therefore, the total complexity is \( O(n(C+d) + \frac{nd^2}{C}) \). During inference, only the previous hidden state needs to be stored, and assuming there are \( h \) heads, the required space is \( O\left(\frac{d^2}{h}\right) \).

  \item \textbf{Time Complexity}. Computing the parameters \( Q, K, V \) for each item requires \( O(nd^2) \) time. The key-value product for each chunk requires \( O(d^2) \) time, and calculating attention within the chunk requires \( O(C^2d) \) time. Thus, the time complexity is \( O(n(C+d)d) \). During inference, the bottleneck lies in the computation of the product between the query vector and the hidden state, with a time complexity of \( O\left(\frac{d^2}{h}\right) \).

\end{itemize}

Since only \( C \) and batch size need to be set to exploit the parallelism of in-chunk computation, \( C \) is a constant independent of \( n \). Thus, our model has \( O(n) \) space-time complexity during parallel training and \( O(1) \) space-time complexity during inference with respect to sequence length $n$. 

\subsection{Relationship with Existing Models}

In this section, we provide a comparative analysis to highlight the architectural distinctions between FuXi-Linear and existing sequence models.

\begin{itemize}[leftmargin=*,align=left]

\item \textbf{FuXi-$\alpha$~\cite{ye2025fuxi} and FuXi-$\beta$~\cite{ye2025fuxib}.} While FuXi-$\alpha$ and FuXi-$\beta$ utilize $O(n^2)$ mechanisms for temporal and positional modeling, FuXi-Linear achieves $O(n)$ complexity by introducing the Linear Positional Channel and Temporal Retention Channel. Beyond the efficiency gain, our model explicitly incorporates periodic temporal patterns, which are not specifically addressed in the previous FuXi variants.

\item \textbf{TiM4Rec~\cite{fan2025tim4rec} and SS4Rec~\cite{xiao2025ss4rec}.} These models incorporate temporal signals by scaling attention weights, a method that often leads to mutual interference between temporal and semantic information and lacks periodicity modeling. In contrast, our Temporal Retention Channel decouples these signals by computing queries and keys derived solely from timestamps. Furthermore, whereas TiM4Rec and SS4Rec rely on the implicit decay of linear attention for order, we introduce the Linear Positional Channel to compensate for the lack of granular positional information in standalone linear mechanisms.

\end{itemize}

%% file: chapters/5_experiments.tex
\section{Experiments}

\subsection{Experiment Setup}\label{ExperimentSetup}

\subsubsection{Datasets}

To evaluate the performance of the proposed FuXi-Linear architecture, we conduct extensive experiments on three public datasets, which are described as follows:

\begin{itemize}[leftmargin=*,align=left]
 \item \textbf{MovieLens-20M} \footnote{https://grouplens.org/datasets/movielens/}: 
    A well-known benchmark dataset for movie recommendations, containing rich user ratings and tagging activities. We use its extensive 20M subset for testing.
  \item \textbf{Kuairand 27k}\footnote{https://kuairand.com/} \cite{gao2022kuairand}: 
    Collected from a video-sharing app, this dataset features long user interaction sequences, capturing high engagement levels.
  \item \textbf{KuaiRec}\footnote{https://kuairec.com/} \cite{gao2022kuairec}: 
    Sourced from another popular video-sharing platform, this dataset also includes long sequences, aiding in the examination of user preferences and engagement.
\end{itemize}

For \textbf{MovieLens-20M}, we adopted exactly the same preprocessing method as HSTU \footnote{https://github.com/facebookresearch/generative-recommenders}. For the other two datasets, a similar preprocessing method was applied. In \textbf{Kuairand-27K}, we retained only the interaction records where users provided no negative feedback and items with more than 100 interactions. The statistical information for the datasets is shown in Table \ref{tab:dataset_statistics}.

\begin{table}[t]
 \caption{Dataset statistics.}
 \centering
 	\setlength{\tabcolsep}{1mm}
 \begin{tabular}{@{} c|c|c|c|c @{}}
 \hline
 \textbf{Dataset} 	  & \textbf{User}   & \textbf{Item} & \textbf{Interactions} & \textbf{Avg. Len.}   \\
 \hline
 MovieLens-20M & 138,493 & 26,744 & 20,000,263 & 144.41   \\
 Kuairand-27K & 27,284 & 131,090 & 97,010,279 & 3555.57 \\ 
 KuaiRec & 7,176 & 9,958 & 12,529,113 & 1745.97 \\ 
 \hline
\end{tabular}
\label{tab:dataset_statistics}
\vspace{-5mm}
\end{table}

\subsubsection{Compared Baseline} 
To comprehensively evaluate the performance of FuXi-Linear, we employ two representative categories of baselines: i) models based on full attention: SASRec \cite{kang2018self}, HSTU \cite{zhai2024actions}, FuXi-$\alpha$ \cite{ye2025fuxi}, FuXi-$\beta$ \cite{ye2025fuxib} ii) models based on linear attention: RecBLR \cite{liu2024behavior}, Mamba4Rec \cite{liu2024mamba4rec}, TTT4Rec \cite{yang2024ttt4rec}, TiM4Rec \cite{fan2025tim4rec}, RetNet \cite{sun2023retentive}. We exclude SS4Rec \cite{xiao2025ss4rec} from our comparison because it exhibits significant numerical instability and incurs prohibitive GPU memory overhead during training.

\subsubsection{Evaluation Metrics}
We utilize the commonly used top-K Hit Ratio (HR@$K$), Normalized Discounted Cumulative Gain (NDCG@$K$, abbreviated as NG@$K$ in tables), and Mean Reciprocal Rank (MRR) to assess recall performance. For each of these metrics, a higher value indicates better performance. By default, we evaluate the ranking of the ground-truth item from the entire set of items and report the performance for $K = 10,50$.

\subsubsection{Implementation}
  We implemented our proposed FuXi-Linear using PyTorch \cite{paszke2019pytorch}. To enable large-scale model training, we utilize multi-card parallelism with the Accelerate library \cite{kotter2012accelerate}. In the following experiment, the position channel of FuXi-Linear was configured with $d_p = 32$, while in the temporal channel, we configured $H_t$ as 8. Based on the distribution of timestamps in the datasets, we empirically set $B$ to 8 for both Kuairand-27K and KuaiRec, and $B$ to 16 for MovieLens-20M.

\subsection{Performance Comparison}
\label{performance_comparsion}

\subsubsection{Parameter Settings}
 For Kuairand-27K, KuaiRec, and ML-20M, we set the number of layers to 4, 4, and 8, respectively, and the embedding size to 128, 128, and 256, respectively. In order to ensure a fair comparison, we adjust the sizes of some hidden layers so that all models are of comparable scale. Detailed parameter settings are listed in Appendix \ref{detailed_parameter_settings}. The sizes of the non-embedding parameters for these models are listed in Table \ref{tab:model_size}.

\begin{table}[htbp]
\centering
\vspace{-5pt}
\caption{The sizes of models across various datasets.}
\vspace{-5pt}
\label{tab:model_size}

\setlength{\tabcolsep}{1mm}{
    \small
    \begin{tabular}{lcccc}
    \toprule
    \textbf{Model} & \textbf{Time-aware} &\textbf{Kuairand 27K} & \textbf{ML-20M} & \textbf{KuaiRec} \\
    \midrule
    SASRec (4d)    & $\times$ & 791.04K & 6.31M & 791.04K \\
    RecBLR (4d)    & $\times$ & 860.67K & 6.85M & 860.67K \\
    Mamba4Rec (4d)      & $\times$ & 1.04M & 7.91M & 1.04M \\
    TTT4Rec (3d)   & $\times$ & 991.12K & 7.38M & 991.12K \\
    RetNet (3d) & $\times$ & 918.66K & 7.34M & 918.66K \\
    TiM4Rec (1d)        & $\checkmark$ & 984.16K & 7.57M & 984.16K \\
    HSTU (4d)      & $\checkmark$ & 926.81K & 7.40M & 926.81K \\
    FuXi-$\alpha$ (2d) &$\checkmark$  & 992.34K & 7.87M & 992.34K \\
    FuXi-$\beta$ (3d)  & $\checkmark$ & 926.30K & 7.35M & 926.30K \\
    FuXi-Linear (1d)      & $\checkmark$ & 917.66K & 7.34M & 917.66K \\
    \bottomrule
    \end{tabular}
}
\end{table}
\vspace{-5pt}

\subsubsection{Public Dataset Performance}

\begin{table*}[h]
\setlength{\abovecaptionskip}{0cm}
\setlength{\belowcaptionskip}{0cm}
\centering
\caption{Overall performance comparison on public datasets.}
\label{tab:performance_comparison}
\setlength{\tabcolsep}{1mm}{
\small
\begin{tabular}{c|c|c|c|c|c|c|c|c|c|c|c|c|c|c|c}
\hline
\textbf{Dataset} & \multicolumn{5}{c|}{\textbf{Kuairand 27K}} & \multicolumn{5}{c|}{\textbf{MovieLens-20M}} & \multicolumn{5}{c}{\textbf{KuaiRec}} \\
\hline
\textbf{Model} & NG@10 & NG@50 & HR@10 & HR@50 & MRR & NG@10 & NG@50 & HR@10 & HR@50 & MRR & NG@10 & NG@50 & HR@10 & HR@50 & MRR \\
\hline \hline
\textbf{SASRec} &
0.0472 & 0.0737 & 0.0902 & 0.2125 & 0.0422 & 
0.1787 & 0.2356 & 0.3138 & 0.5713 & 0.1524 &
0.1154 & 0.1612 & 0.1895 & 0.4027 & 0.1059 \\

\textbf{RecBLR} &
0.0451 & 0.0711 & 0.0849 & 0.2049 & 0.0409 &
0.1739 & 0.2316 & 0.3049 & 0.5659 & 0.1491 &
0.1102 & 0.1565 & 0.1837 & 0.3983 & 0.1010 \\

\textbf{Mamba4Rec} & 
0.0470 & 0.0743 & 0.0881 & 0.2140 & 0.0427 &
0.1862 & 0.2426 & 0.3220 & 0.5765 & 0.1595 &
0.1165 & 0.1622 & 0.1951 & 0.4068 & 0.1058 \\

\textbf{TTT4Rec} &
0.0467 & 0.0743 & 0.0874 & 0.2148 & 0.0425 &
0.1722 & 0.2277 & 0.3013 & 0.5526 & 0.1475 &
0.1135 & 0.1622 & 0.1885 & 0.4151 & 0.1042 \\

\textbf{RetNet} &
0.0448 & 0.0715 & 0.0827 & 0.2061 & 0.0413 &
0.1877 & 0.2445 & 0.3236 & 0.5803 & 0.1610 &
0.1143 & 0.1618 & 0.1915 & 0.4115 & 0.1044 \\

\textbf{TiM4Rec} &
0.0454 & 0.0720 & 0.0850 & 0.2076 & 0.0414 &
0.1850 & 0.2418 & 0.3206 & 0.5772 & 0.1584 &
0.1149 & 0.1609 & 0.1894 & 0.4029 & 0.1053 \\

\textbf{HSTU} &
0.0534 & 0.0822 & 0.0994 & \underline{0.2319} & 0.0479 &
0.2095 & 0.2663 & 0.3548 & 0.6107 & 0.1798 &
0.1265 & 0.1755 & 0.2074 & \underline{0.4358} & 0.1155 \\

\textbf{FuXi-$\alpha$} &
\underline{0.0543} & \underline{0.0825} & \underline{0.1005} & 0.2300 & \underline{0.0487} &
0.2102 & 0.2671 & 0.3550 & 0.6121 & 0.1806 & 
\underline{0.1281} & \underline{0.1764} & \underline{0.2090} & 0.4326 & \underline{0.1170} \\

\textbf{FuXi-$\beta$} &
0.0526 & 0.0809 &  0.0974 & 0.2279 & 0.0472 & 
\underline{0.2116} & \underline{0.2688} & \underline{0.3562} & \underline{0.6146} & \underline{0.1820} & 
0.1238 & 0.1702 & 0.2039 & 0.4197 & 0.1125  \\

\textbf{FuXi-Linear} &
\textbf{0.0609} & \textbf{0.0906} & \textbf{0.1124} & \textbf{0.2488} & \textbf{0.0540} & 
\textbf{0.2131} & \textbf{0.2700} & \textbf{0.3592} & \textbf{0.6161} & \textbf{0.1830} &
\textbf{0.1368} & \textbf{0.1851} & \textbf{0.2242} & \textbf{0.4486} & \textbf{0.1235}\\
\hline
\end{tabular}
}
\end{table*}

In this section, we evaluate the effectiveness of \textsf{FuXi-Linear} by comparing it against state-of-the-art baselines across three public datasets. The empirical results, summarized in Table \ref{tab:performance_comparison}, lead to the following key observations:
\begin{itemize}[leftmargin=*,align=left]
    \item FuXi-Linear consistently achieves the best performance across all evaluation metrics, with improvements being particularly substantial in long-sequence scenarios. Specifically, on the Kuairand-27K and KuaiRec datasets, FuXi-Linear significantly advances the state-of-the-art by achieving average relative improvements of 9.26\%, 7.24\%, 9.01\%, 5.11\%, and 8.33\% in NDCG@10, NDCG@50, HR@10, HR@50, and MRR, respectively.

    \item The experimental results underscore that the modeling of temporal signals is essential for recommendations. Models that effectively integrate temporal information, such as HSTU, FuXi-alpha, FuXi-beta, and FuXi-Linear, consistently and significantly outperform time-agnostic models like SASRec and Mamba4Rec. 

    \item While temporal information is crucial, the efficacy of its utilization is equally important for model performance. For instance, although TiM4Rec attempts to incorporate temporal signals, it fails to achieve competitive results. This is likely due to its overly complex transformations, which lead to the loss of raw temporal patterns and a failure to harmonize time with semantic features. As a result, its performance is similar to that of linear models that ignore temporal information.
\end{itemize}

\subsection{Efficiency Comparison}

\begin{figure*}[ht]
    \centering
    \subfigure[Prefill]{
        \includegraphics[width=0.48\linewidth]{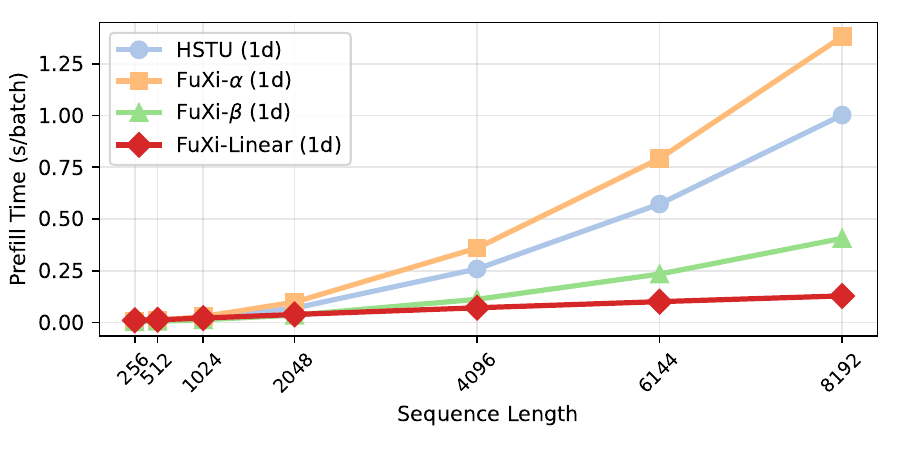}
        \label{fig:efficiency-prefill}
    }
    \subfigure[Decode]{
        \includegraphics[width=0.48\linewidth]{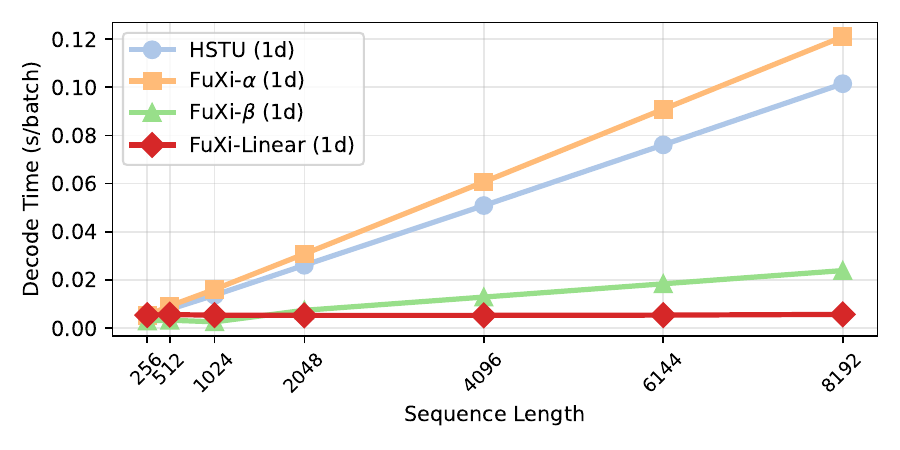}
        \label{fig:efficiency-decode}
    }
    \vspace{-5pt}
    \caption{The efficiency comparison across different sequence lengths.}
    \vspace{-5pt}
    \label{fig:efficiency}
\end{figure*}

We select several of the best-performing models—FuXi-Linear, FuXi-$\alpha$, FuXi-$\beta$, and HSTU—for a comprehensive efficiency comparison. We evaluate their performance under two scenarios:
\begin{itemize}[leftmargin=*,align=left]
   \item \textbf{Prefill}: Computes cache information based on existing user interaction sequences.
   \item \textbf{Decode}: Generating the next item and updating the cache to provide real-time recommendations.
\end{itemize}
For a fair evaluation, all models are configured with 2 layers ($L=2$), an embedding size of $d=256$, and consistent head dimensions. We set $d_{\text{FFN}} = d$, and for FuXi-Linear, the chunk size is $C=128$. To more objectively reflect the intrinsic computational complexity of each model, we employ FlashAttention-2 \cite{dao2023flashattention} to accelerate all models.
The experiments use batch sizes of 64 for prefilling and 1024 for decoding, with results averaged over 10 batches. As illustrated in Figure \ref{fig:efficiency}, FuXi-Linear demonstrates a substantial speedup over other models as the sequence length increases, owing to its lower computational complexity. At a sequence length of 8k, FuXi-Linear outperforms FuXi-$\alpha$, FuXi-$\beta$, and HSTU in the Prefill stage with speedup ratios of 10$\times$, 3.1$\times$, and 7.8$\times$, respectively. In the Decode stage, these ratios reach 21$\times$, 4.2$\times$, and 18$\times$. These results highlight that FuXi-Linear offers a highly efficient solution for processing ultra-long sequences in recommendation systems.

\subsection{Different Variants Comparison}
\subsubsection{Comparison of Different Temporal Encoding Methods.}

To evaluate the effectiveness of the Linear Temporal Channel, we conduct a comparative analysis against several representative temporal modeling methods. The results on the Kuairand-27K dataset are summarized in Table \ref{tab:temporal_comparison}. The empirical results indicate that while absolute sinusoidal encoding \cite{vaswani2017attention} is computationally efficient, it yields limited performance gains as it fails to capture complex temporal dependencies. Although TiSSD \cite{fan2025tim4rec} also has linear complexity, it fails to effectively disentangle semantic and temporal signals, resulting in a slight decline in performance. Furthermore, although models such as HSTU \cite{zhai2024actions} achieve better accuracy by leveraging relative temporal information, they are hindered by quadratic complexity. In contrast, our proposed method outperforms all baselines by effectively modeling both relative temporal intervals and periodic patterns, while maintaining linear complexity.

\begin{table}[htbp]
    \centering
    \vspace{-5pt}
    \caption{Comparison of different temporal encoding methods on Kuairand-27K.}
    \vspace{-5pt}
    \label{tab:temporal_comparison}
    
    \setlength{\tabcolsep}{0.5mm}{
    \footnotesize
    \begin{tabular}{lcccccc}
        \toprule
        \textbf{Method} & \textbf{NG@10} & \textbf{NG@50} & \textbf{HR@10} & \textbf{HR@50} & \textbf{MRR} & \textbf{Complexity} \\
        \midrule
        Ours & \textbf{0.0609} & \textbf{0.0906} & \textbf{0.1124} & \textbf{0.2488} & \textbf{0.0540} & $O(nd)$ \\
        Method in FuXi-$\alpha$     & 0.0564 & 0.0858 & 0.1045 & 0.2397 & 0.0504 & $O(n^2 d)$ \\
        Method in FuXi-$\beta$      & 0.0540 & 0.0831 & 0.0996 & 0.2327 & 0.0487 & $O(n^2 d)$ \\
        Bias in HSTU                & 0.0538 & 0.0825 & 0.1007 & 0.2326 & 0.0480 & $O(n^2)$ \\
        Absolute Embedding          & 0.0504 & 0.0776 & 0.0948 & 0.2199 & 0.0452& $O(nd)$ \\
        TiSSD (in TiM4Rec)          & 0.0468 & 0.0737 & 0.0887 & 0.2132 & 0.0421 & $O(nd)$ \\
        None                        & 0.0471 & 0.0738 & 0.0892 & 0.2122 & 0.0424 & 0 \\
        \bottomrule
    \end{tabular}
    }
\end{table}

\subsubsection{Comparison of Different Positional Encoding Methods.}
To evaluate the Linear Positional Channel, we conduct a comparative study with several widely-used positional encoding schemes, including relative position-based methods such as FuXi \cite{ye2025fuxi,ye2025fuxib}, T5 \cite{raffel2020exploring}, and Alibi \cite{press2021train} as well as RoPE \cite{su2024roformer}. The results are summarized in Table \ref{tab:positional_encoding}. The empirical results demonstrate that our proposed method achieves competitive performance compared to the strongest relative positional modeling baselines. Notably, while high-performing relative methods often incur quadratic computational overhead, our approach delivers comparable effectiveness while maintaining linear complexity, demonstrating a superior trade-off between modeling capacity and efficiency.

\begin{table}[htbp]
    \centering
    \vspace{-5pt}
    \caption{Comparison of different positional encoding methods on Kuairand-27K.}
    \vspace{-5pt}
    \label{tab:positional_encoding}
    \setlength{\tabcolsep}{0.9mm}{
    \footnotesize
    \begin{tabular}{lcccccc}
        \toprule
        Method & NG@10 & NG@50 & HR@10 & HR@50 & MRR & Additional Cost \\
        \midrule
        Ours & \textbf{0.0609} & \textbf{0.0906} & \textbf{0.1124} & \textbf{0.2488} & \textbf{0.0540} & $O(nd)$ \\
        Used in FuXi-$\alpha$/$\beta$ & 0.0607 & 0.0900 & 0.1115 & 0.2456 & 0.0539 & $O(n^2d)$ \\
        RoPE & 0.0575 & 0.0872 & 0.1071 & 0.2433 & 0.0511 & $O(nd)$ \\
        T5Bias & 0.0580 & 0.0866 & 0.1077 & 0.2389 & 0.0513 & $O(n^2)$ \\
        Alibi & 0.0574 & 0.0865 & 0.1065 & 0.2407 & 0.0509 & $O(n^2)$ \\
        None & 0.0568 & 0.0854 & 0.1067 & 0.2385 & 0.0501 & 0 \\
        \bottomrule
    \end{tabular}
    }
\end{table}


\subsection{Ablation Study}\label{ablation_study}

We conduct an ablation study on the Kuairand-27K dataset to evaluate the contribution of each component within FuXi-Linear, with results summarized in Table \ref{tab:ablation_study}.

The empirical evidence shows that removing any single module results in a performance decline, highlighting their collective necessity. Notably, the temporal channel proves most critical; its removal leads to the most substantial performance drop. Specifically, excluding $Q_t$ and $K_t$ impairs the model's capacity to capture periodic information, which is vital for temporal modeling. The positional channel ranks second in importance; while temporal sequences offer coarse ordering, the positional channel provides the fine-grained structural information necessary to resolve these limitations. Finally, although the retention channel yields the smallest marginal improvement, it enhances the model's ability to selectively retrieve relevant historical data through semantic query-key attention.

\begin{table}[htbp]
    \centering
    \vspace{-5pt}
    \caption{Ablation Study of FuXi-Linear on Kuairand-27K.}
    \vspace{-5pt}
    \label{tab:ablation_study}
    
    \setlength{\tabcolsep}{0.8mm}{
    \small
    \begin{tabular}{lccccc}
        \toprule
        \textbf{Setting} & \textbf{NG@10} & \textbf{NG@50} & \textbf{HR@10} & \textbf{HR@50} & \textbf{MRR} \\
        \midrule
        Base                  & \textbf{0.0609} & \textbf{0.0906} & \textbf{0.1124} & \textbf{0.2488} & \textbf{0.0540} \\
        w/o temporal channel & 0.0471 & 0.0738 & 0.0892 & 0.2122 & 0.0424 \\
        w/o $Q_t, K_t$ & 0.0566 & 0.0857 & 0.1051 & 0.2388 & 0.0504 \\
        w/o positional channel & 0.0568 & 0.0854 & 0.1067 & 0.2385 & 0.0501 \\
        w/o retention channel & 0.0588 & 0.0880 & 0.1071 & 0.2409 & 0.0527 \\
        \bottomrule
    \end{tabular}
    }
\end{table}

\textbf{Hyper-parameter Study.} Complementing the module-level ablation, we further investigate the sensitivity of FuXi-Linear to its core internal parameters, including the number of temporal heads ($H_t$), retention heads ($H$), and the positional embedding dimension ($d_p$). These analyses reinforce our design intuitions, such as the necessity of multi-scale temporal modeling and the effectiveness of the multi-head mechanism in reducing state overhead. Detailed results are provided in Appendix \ref{hyper_study}.

\subsection{Scaling Experiment}

\begin{figure}
    \centering
    \setlength{\belowcaptionskip}{-10pt}
        \includegraphics[width=0.95\linewidth]{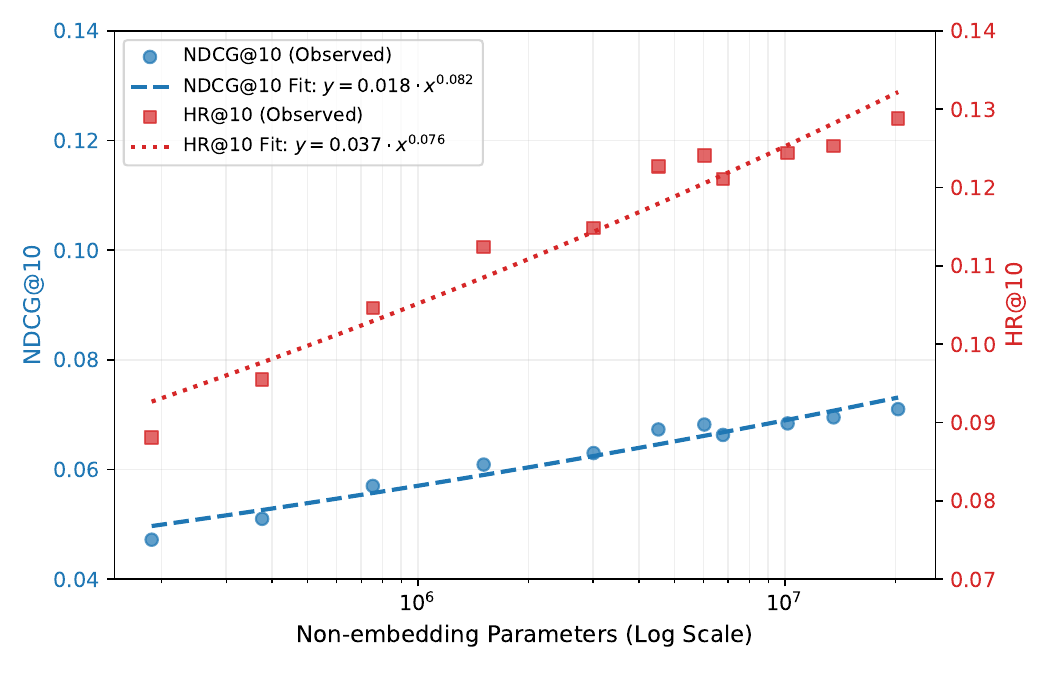}
    \caption{Scaling performance of FuXi-Linear with respect to model size on Kuairand-27K.}
    \label{fig:scaling}
    \vspace{-5pt}
\end{figure}

To verify the scalability of our model, we conducted experiments using the Kuairand-27K dataset. We fixed the sequence length at $n = 1024$ and set the head dimension to 32 with $d_{\text{FFN}} = 4d$. We then incrementally enlarged the model by adjusting only the embedding dimension $d$ and the number of layers $L$. As shown in Figure \ref{fig:scaling}, FuXi-Linear exhibits a robust power-law scaling property across multiple metrics as non-embedding parameters increase. Specifically, NG@10 and HR@10 scale consistently from 0.0472 and 0.0881 (188K parameters) to 0.0710 and 0.1288 (20M parameters), respectively. This sustained improvement across two orders of magnitude in model size underscores the excellent capacity and stability of the FuXi-Linear architecture.

%% file: chapters/6_conclusions.tex
\section{Conclusion}
In this work, we presented FuXi-Linear, a novel linear-complexity architecture designed to resolve the efficiency-effectiveness dilemma in long-term sequential recommendation. We introduced the Temporal Retention Channel, which effectively captures periodic patterns while decoupling temporal signals from semantic features to prevent mutual interference. Furthermore, we developed the Linear Positional Channel, which leverages learnable kernels to approximate the expressive power of relative position encoding within a strictly linear-time recurrent formulation. Extensive experiments on three real-world benchmarks demonstrate that FuXi-Linear significantly outperforms state-of-the-art baselines, particularly in long-sequence scenarios. We further validated that our architecture possesses a robust power-law scaling property at a thousand-length scale, demonstrating sustained performance gains as model capacity increases. Compared to competitive baselines, FuXi-Linear achieves up to 10$\times$ speedup in the prefill stage and 21$\times$ higher inference throughput during decoding. Overall, FuXi-Linear provides a promising and scalable solution for ultra-long sequence modeling in modern recommender systems. In the future, we plan to further optimize the efficiency of our model and extend our method to encompass more diverse and complex multi-behavior user sequences.

%% file: chapters/7_appendix.tex
\appendix
\balance

\section{Detailed Parameter Settings}\label{detailed_parameter_settings}

By default, we set the number of training epochs to 100 for all datasets. For Kuairand-27K and KuaiRec, we set the batch size to 128 and the learning rate to $10^{-3}$. For MovieLens-20M, we use a batch size of 1024 and a learning rate of $4 \times 10^{-3}$. However, due to the unstable training of architectures such as SASRec and Mamba at a learning rate of $4 \times 10^{-3}$, which makes effective optimization challenging, we used a learning rate of $2 \times 10^{-4}$ for these models, increasing the number of training epochs to 200.

For models such as SASRec, FuXi-$\alpha$, FuXi-$\beta$, RetNet, TTT4Rec, and RecBLR, we increased model size by adjusting the hidden layer parameters of the feed-forward network. Let $d$ denote the embedding size; we set the hidden layer parameters to $4d$, $2d$, $3d$, $3d$, $3d$, and $4d$, respectively. For HSTU, we increased model size by adjusting the dimension of the value vector in the attention mechanism to $4d$. For Mamba and TiM4Rec, we set the expand parameter to 2, $d_{\text{conv}}$ to 4, and ensure $d_{\text{state}}$ matches the head dimension of query-key attention based models.

\begin{figure*}[b]
    \centering
    \subfigure[Number of temporal heads]{
        \includegraphics[width=0.32\linewidth]{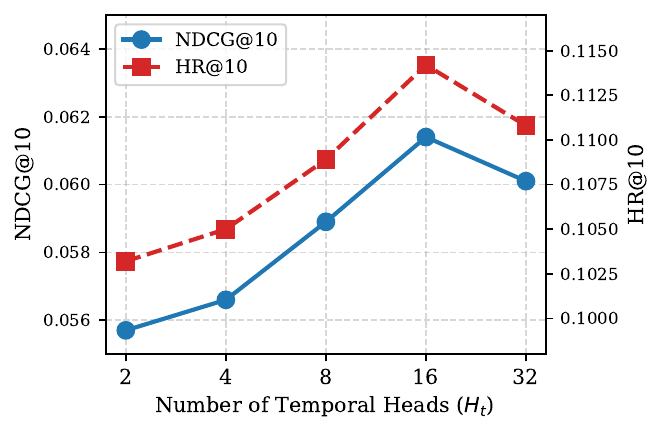}
        \label{fig:hyper-h-t}
    }
    \subfigure[Number of retention heads]{
        \includegraphics[width=0.32\linewidth]{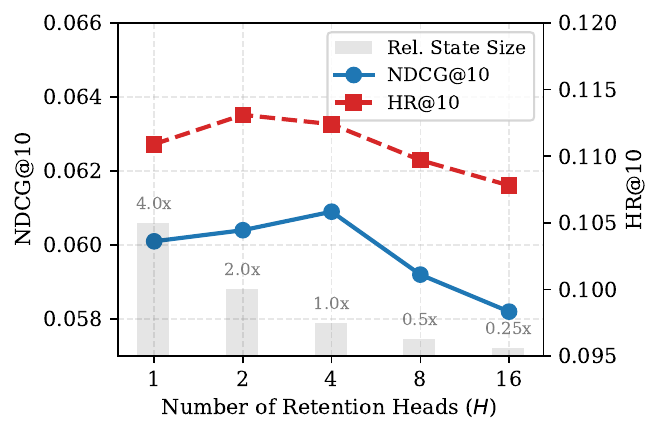}
        \label{fig:hyper-h-ret}
    }
    \subfigure[Positional embedding dimension]{
        \includegraphics[width=0.32\linewidth]{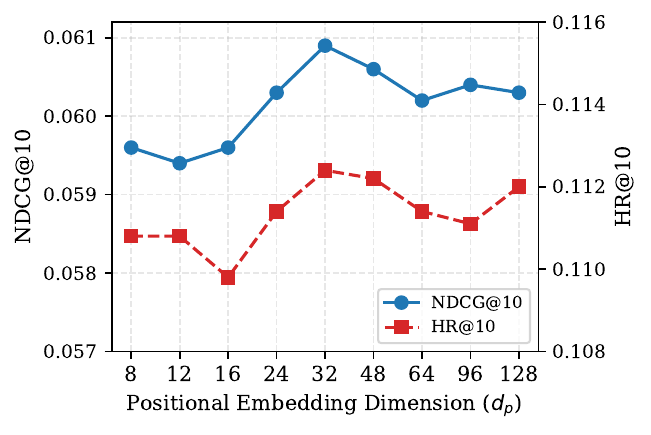}
        \label{fig:hyper-d-p}
    }
    \caption{Hyperparameter study on FuXi-Linear performance.}
    \label{fig:hyperparameters}
\end{figure*}

\section{Hyperparameter Study}\label{hyper_study}

\subsection{Number of Temporal Heads ($H_t$)}
We vary $H_t$ while fixing the total temporal coverage ($B^{H_t}=32$) to examine the model's multi-scale temporal modeling capacity. As shown in Figure \ref{fig:hyper-h-t}, performance improves consistently as $H_t$ increases from 2 to 16. This trend validates that providing a rich set of discrete time-scales is essential for the model to perceive and aggregate the vast range of temporal dynamics present in long-term recommendation sequences.  However, performance declines at $H_t = 32$ due to a representational bottleneck: with a fixed total dimension, an excessive number of heads results in prohibitively small dimensions for the value vectors $v_t$, impairing the expressive power of temporal features at each scale.

\subsection{Number of Retention Heads ($H$)}
We vary the number of retention heads $H$ with a fixed dimension $d=128$, where the total recurrent state size for inference is $d^2/H$ (Section \ref{Analysis}). Figure~\ref{fig:hyper-h-ret} shows that performance peaks at $H=4$. Small $H$ (1 or 2) yields suboptimal results despite a larger state capacity, as fewer heads limit the model's ability to decouple diverse user interests. Conversely, when $H > 4$, the shrinking head dimensions and overall state size lead to information loss, causing performance to degrade. These results demonstrate that FuXi-Linear achieves superior performance without necessitating a large state size.

\subsection{Positional Dimension ($d_p$)}
We vary the embedding dimension of the positional channel $d_p$ from 8 to 128 to characterize the learnable kernel's expressive power. As illustrated in Figure~\ref{fig:hyper-d-p}, performance rises sharply up to $d_p=32$, indicating that low dimensions lack sufficient rank to approximate complex positional dependencies. Notably, performance enters a stable plateau beyond $d_p=32$ with only marginal fluctuations. This suggests that a relatively low dimension ($d_p=32$) is sufficient for effective positional modeling.